\documentclass[fleqn,12pt,twoside]{article}
\usepackage{graphicx}
\begin{document}

\title{$\Lambda$ and $\phi$ Production in Heavy Ion Collisions}
\author{Andr\'e Mischke \it{for the NA49 collaboration}\\
\small{Gesellschaft f\"ur Schwerionenforschung, Darmstadt, Germany}\\
\small{present address: NIKHEF and University of Utrecht, The Netherlands}}  
\maketitle

\begin{abstract}
Recent NA49 results on $\Lambda$ and $\phi$ production in central Pb-Pb 
\mbox{collisions} at beam energies from 40 to 158~A$\cdot$GeV are presented.
\end{abstract}

The measurement of strange baryons like $\Lambda$(uds), which
contain between 30 and 60$\%$ of the total strangeness produced, 
allows to study simultaneously strangeness
production and the effect of baryon density in A-A collisions.
Essentially half of the $\bar{\rm s}$ quarks are contained in K$^+$.
The $\phi$ meson consists of a s$\bar{\rm s}$ pair,
and should therefore be more sensitive than kaons and lambdas to the
production mechanism in the early stage of the collision.

Since 1994 the NA49 collaboration has investigated hadron production
in central Pb-Pb collisions at 158~A$\cdot$GeV over a large range 
of rapidity and transverse momentum.
Within the framework of the NA49 energy scan programme,
started five years later, large data sets at lower energies
(20, 30, 40, 80~A$\cdot$GeV) have been recorded, for details see 
Ref.~\cite{Marek03}.

In the following, results on $\Lambda$ hyperon and $\phi$
meson production in central Pb-Pb collisions at beam energies
40-158~A$\cdot$GeV are presented. Starting from the reconstruction 
method we will focus on the energy dependence of the inverse slope
parameter and the particle multiplicities.\\
\begin{figure}[t]
 \resizebox{0.9\textwidth}{!}{%
 \includegraphics{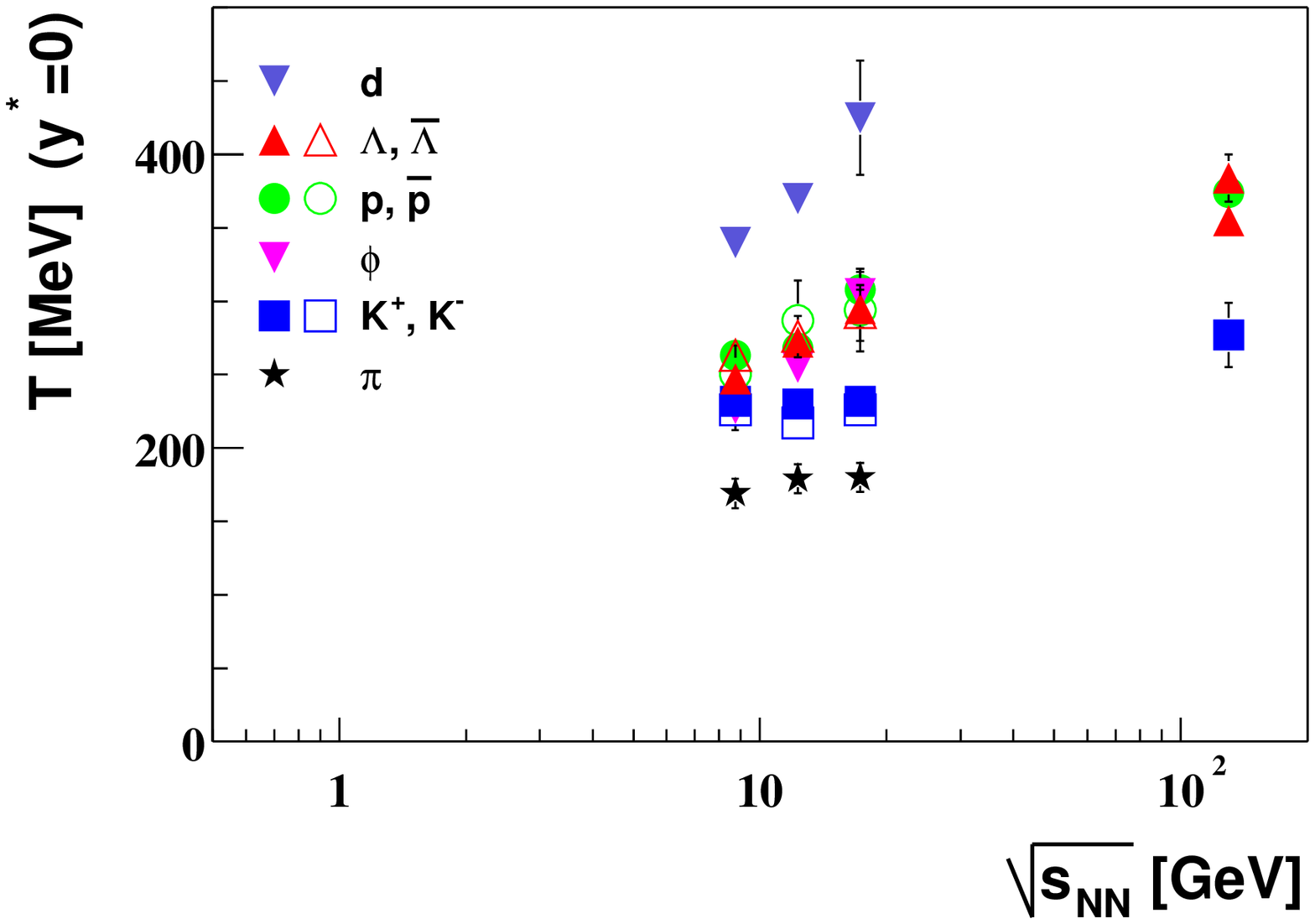}
  \hspace*{3cm}
  \includegraphics{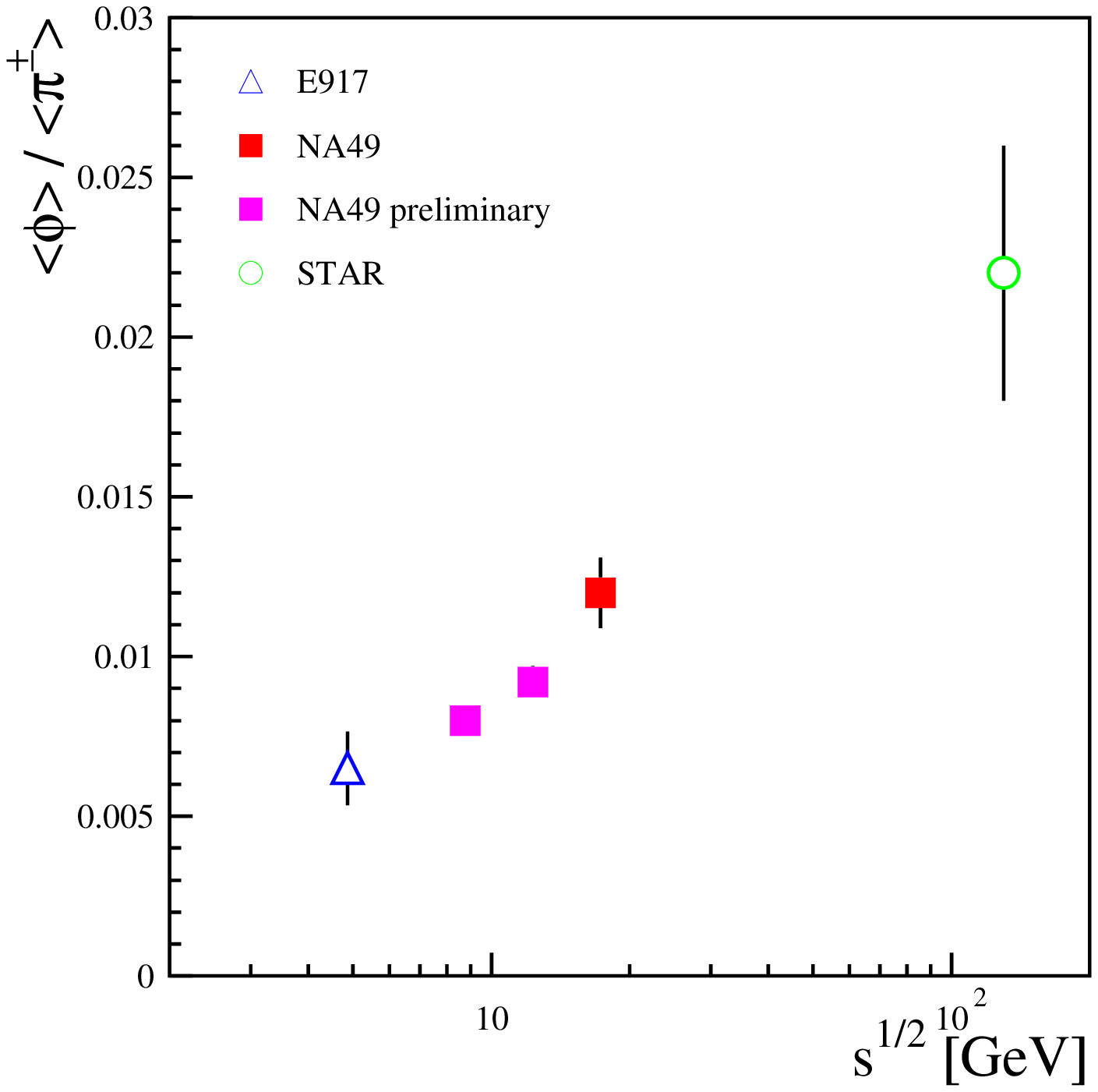}
 }
  \vspace{-0.5cm}
\caption{\protect \footnotesize
The slope parameters T for different particles (left) and the 
$\phi$ multiplicity ratios (right) as function of energy.}
\label{fig:1}
\end{figure}
\begin{figure*}
 \resizebox{0.9\textwidth}{!}{%
  \includegraphics{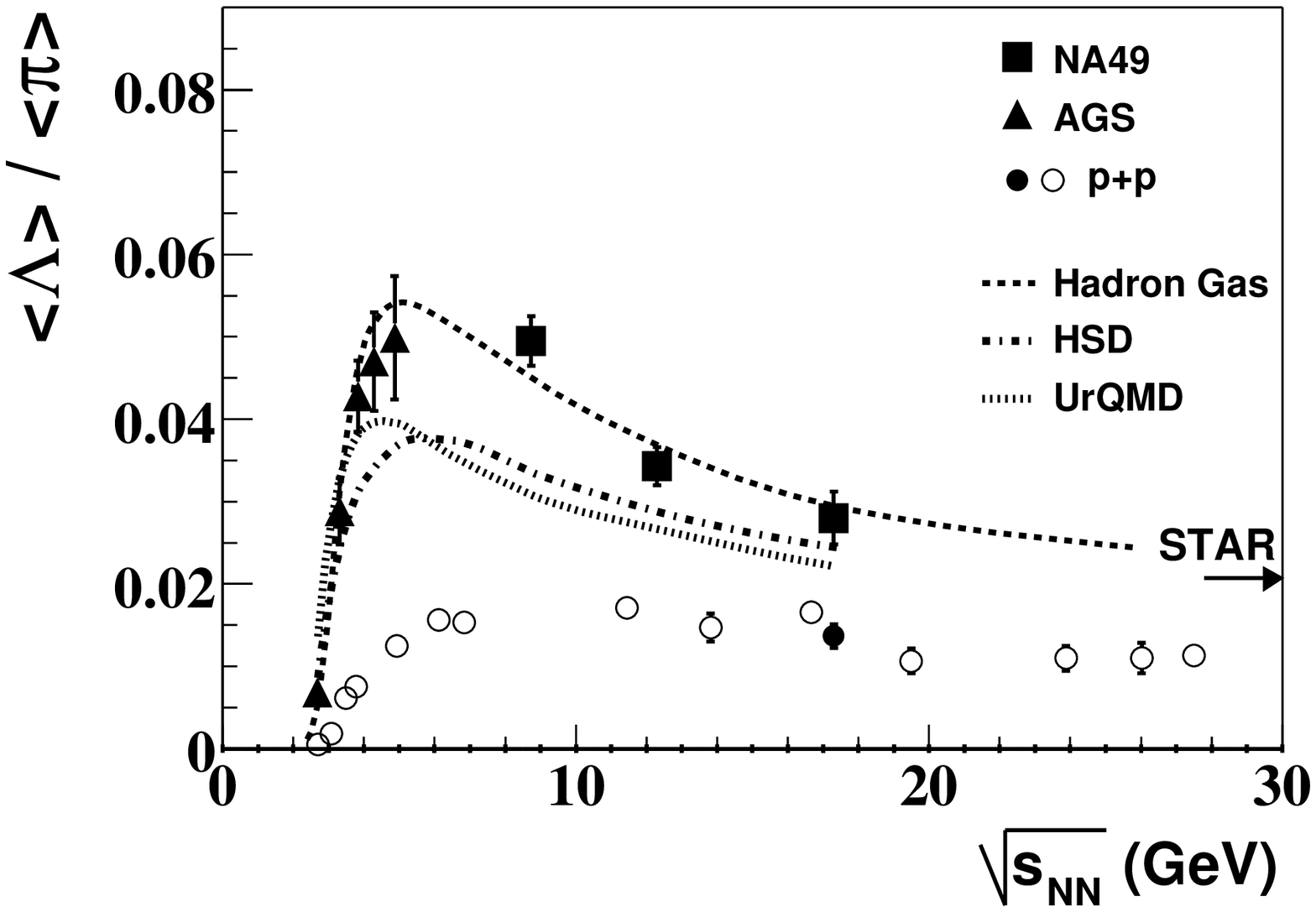}
  \hspace*{4cm}
  \includegraphics{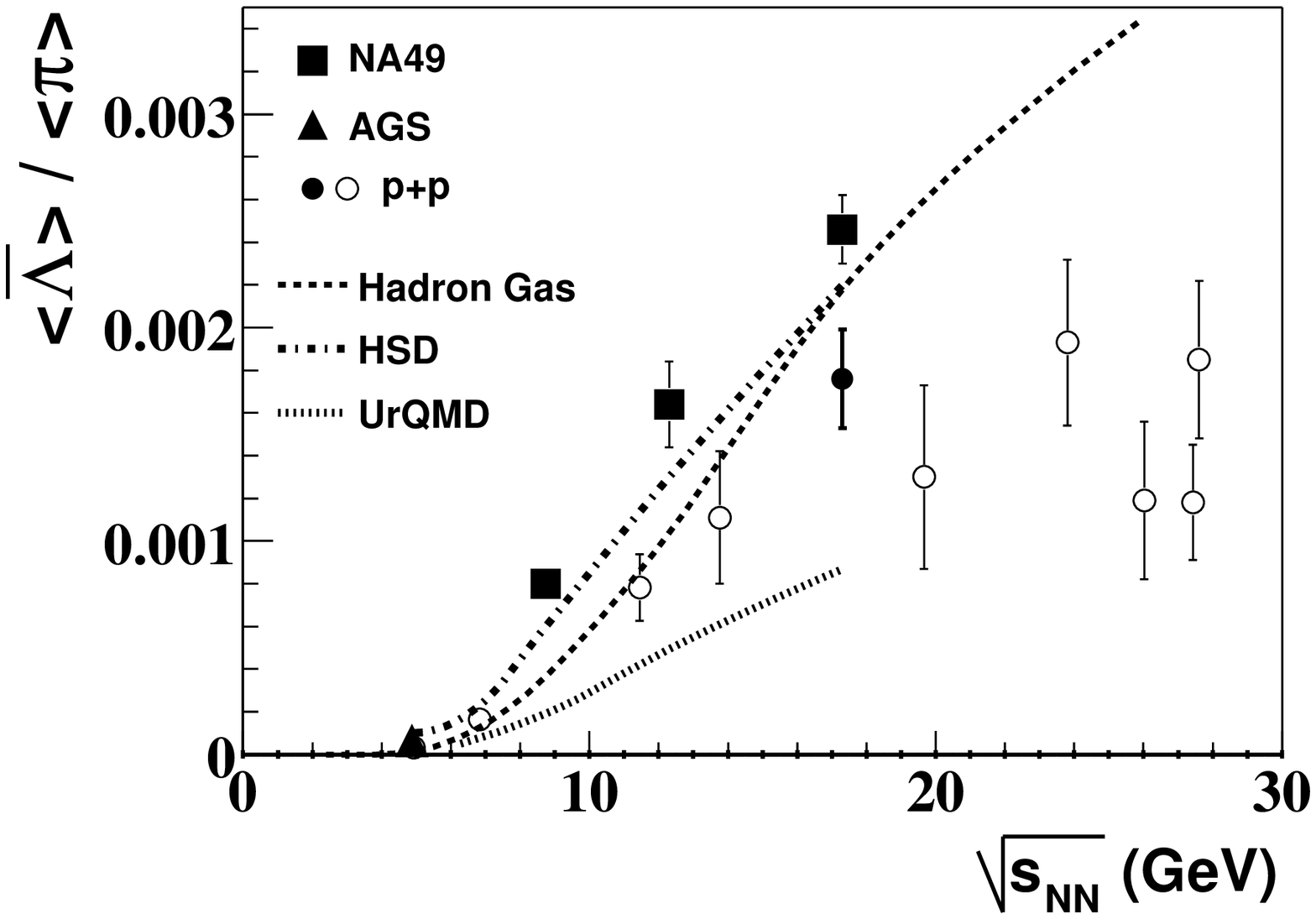}
 }
 \vspace{-0.5cm}
\caption{\protect \footnotesize
Energy dependence of $\Lambda$-to-pion ratio in central Pb-Pb (Au-Au)
and p-p (open symbols) collisions.}
\label{fig:2}
\end{figure*}

In NA49, neutral strange and multi-strange particles are identified
by the topology of the charged decay products, which are measured with 
the TPCs.
The invariant mass distributions of $\Lambda$ and $\bar{\Lambda}$
hyperons are given in Ref.~\cite{Mis03}. The mass resolution
($\sigma_{\rm m}$) of 2~MeV/$c^{2}$ is remarkably good.
After corrections for acceptance and reconstruction efficiency the
$m_{\rm T}$ spectra and rapidity distributions were obtained
from the raw data yields~\cite{Mis03}.
The left plot of Fig.\ref{fig:1} shows the energy dependence of the
inverse slope parameter $T$ of $\Lambda$ and $\overline{\Lambda}$,
which exhibits a slight increase. This increase is also observed
for protons, $\phi$ and deuterons. 
In contrast, the $\pi$ and K slope parameter stays constant.

The total $\Lambda$ and $\overline{\Lambda}$ yields per event
were obtained by integration of the distributions over rapidity and
$p_{\rm T}$ with only small extrapolations into unmeasured regions.
The $\Lambda/\pi$ and $\overline{\Lambda}/\pi$ ratio 
from A-A and p-p collisions as a function of cms energy 
$\sqrt{s_{\rm NN}}$ is shown in Fig.\ref{fig:2}, where 
$\pi=1.5(\pi^++\pi^-)$.
%
%
The $\Lambda/\pi$ ratio steeply increases at AGS energies, reaches
a maximum and drops at SPS energies.
Since ${\rm K}^{+}$ carry the major fraction of the produced
$\bar{\rm s}$ quarks one expect the ${\rm K}^{+}/\pi^{+}$ ratio to
show a similar behavior as the $\Lambda/\pi$ ratio (using
strangeness conservation) which is indeed the case~\cite{Marek03}.

The enhancement of strangeness production in heavy ion collisions
compared to p-p is not a unique signature for the deconfined state,
since this enhancement is at low AGS energies, where a phase transition
is not expected, about seven times higher than at top SPS 
energies~\cite{MRTqm02}.
Instead, rescattering processes like associated production
$\pi$N $\rightarrow \Lambda$K play an important role at lower energies.
          
In comparison to the energy dependence of $\Lambda$ production, the
$\overline{\Lambda}/\pi$ ratio shows a monotonic increase similar
to the K$^-/\pi^-$ ratio~\cite{Marek03}.
The measurements at 20 and 30~A$\cdot$GeV will clarify
whether there is also a structure like for the K$^-/\pi^-$ ratio.
The differences in the excitation function of $\Lambda$ and
$\overline{\Lambda}$ can be attributed to their different production
mechanisms and the effect of net-baryon density.

The transport models UrQMD and HSD as well as the statistical model 
reflect the main
trend of the energy dependence of the $\Lambda/\pi$ and
$\overline{\Lambda}/\pi$ ratio, except the UrQMD model for the
$\overline{\Lambda}/\pi$ ratio. Since the lambda rapidity distributions
are well described by the transport models the discrepancies can
be attributed to the over-prediction of the pion multiplicities.\\

The $\phi$ meson is measured in NA49 via the invariant mass of its decay
products K$^+$K$^-$~\cite{Fri03}. The combinatorial background from
random pairs is well described by means of the event-mixing method.
The invariant mass distributions and the obtained rapidity spectra
for 40, 80 and 158~A$\cdot$GeV are given in Refs.~\cite{Fri03}.
The extracted total $\phi$ yields normalized to the average number of pions
$\pi^{\pm}=0.5(\pi^++\pi^-)$ are illustrated in the right plot of
Fig.\ref{fig:1}.
This ratio shows the same monotonic rise from AGS to RHIC energies
as the K$^-/\pi^-$ ratio.\\

In summary, the $\Lambda/\pi$ ratio shows a maximum around 30~A$\cdot$GeV, 
whereas the $\overline{\Lambda}/\pi$ ratio exhibits a continuous rise.
The $\phi$/$\pi$ ratio also increases monotonically with energy.
In the measured energy range the inverse slope parameter slightly increase
for $\Lambda$, protons, $\phi$, and deuteron, except for $\pi$ and K.

\end{document}